\documentclass[journal]{IEEEtran}

\usepackage{amsmath,amsfonts}
\usepackage{amssymb}
\usepackage{array}
\usepackage{graphicx}
\usepackage[hyphens]{url}
\usepackage{cite}
\usepackage{stfloats}
\usepackage{textcomp}
\usepackage{verbatim}
\usepackage[caption=false,font=footnotesize]{subfig}

\usepackage{algorithm}
\usepackage{algpseudocodex}

\usepackage{tabularx}
\usepackage{multirow}
\usepackage{booktabs}
\usepackage{makecell}
\usepackage{threeparttable}

\usepackage[colorlinks=true, linkcolor=blue, citecolor=blue, urlcolor=blue]{hyperref}

\newtheorem{proposition}{Proposition}

\newcommand{\nomlabels}{\IEEEusemathlabelsep\IEEEsetlabelwidth{$\cdot(b),\cdot(z)$}}

\begin{document}

\title{\fontsize{24pt}{28.8pt}\selectfont Valuing Static Synchronous Series Compensators in the U.S. Grid: Planning with Transmission Feedback}

\author{Wei Ai,~\IEEEmembership{Student Member,~IEEE},
        Vladimir Dvorkin,~\IEEEmembership{Member,~IEEE},
        Michael T. Craig
\thanks{All authors are with the University of Michigan, Ann Arbor, MI 48109 USA: W. Ai and M. T. Craig with the School for Environment and Sustainability, W. Ai and V. Dvorkin with the Department of Electrical Engineering and Computer Science, and M. T. Craig also with the Department of Industrial and Operations Engineering. Corresponding author: W. Ai (email: weiai@umich.edu). W. Ai and M. T. Craig thank the U.S. National Science Foundation for funding under Grant \#2142421.}%
}

\maketitle

\begin{abstract}
Flexible AC Transmission Systems, particularly Static Synchronous Series Compensators (SSSCs), can increase the transfer capability of existing networks and complement constrained transmission expansion, but their value, optimal scale, and best locations in U.S. power system planning remain unquantified. This paper develops a capacity expansion model of the contiguous U.S. power system for 2050, incorporating SSSC-modified power flow equations. Expanding a line by adding parallel circuits lowers its equivalent impedance and, for a given SSSC rating, the series voltage the SSSC can inject. Through this transmission feedback, line flows depend nonlinearly on capacity decisions, making the planning problem nonconvex. We solve it by sequential linear programming (SLP) with adaptive proximal regularization. We prove the convergence to a Karush--Kuhn--Tucker point, which does not hold for the conventional fixed-impedance iteration. We also demonstrate that SLP converges faster to lower-cost plans. In the baseline scenario, the least-cost plan deploys 291~GVAr of SSSCs across the contiguous U.S. SSSCs reduce annualized system cost by \$6.75 billion (2.1\%) with an overall benefit--cost ratio (BCR) of 6.4 and lower the average wholesale electricity price by 5.1\%. Alternatively, SSSCs could avoid about 75\% of the transmission expansion without increasing system cost. Most of the economic value of SSSCs could be delivered by deployment in high-value locations: the highest-value 1~GVAr, located mainly near Eastern U.S. load centers, yields a BCR above 100, and one-tenth of the optimal SSSC capacity captures half of the savings. The overall BCR of SSSCs ranges from 4.1 to 9.1 across the demand, decarbonization, and device cost scenarios considered.
\end{abstract}

\begin{IEEEkeywords}
FACTS, grid-enhancing technologies, static synchronous series compensator, power system planning, transmission expansion, impedance feedback
\end{IEEEkeywords}

\section*{Nomenclature}
Indices are omitted below for brevity. Bold symbols denote vectors, and overlines denote upper bounds.
\subsection*{Sets}
\begin{IEEEdescription}[\nomlabels]
\item[$\mathcal{B},b$] Set and index of buses.
\item[$\mathcal{G},g$] Set and index of generators and electrolyzers.
\item[$\mathcal{I},i$] Set and index of DC lines.
\item[$\mathcal{J},j$] Set and index of independent cycles in the AC network.
\item[$\mathcal{K},k$] Set and index of periods.
\item[$\mathcal{L},\ell$] Set and index of AC transmission lines (branches of the reduced network).
\item[$\mathcal{M},m$] Set and index of AC loss segments.
\item[$\mathcal{S},s$] Set and index of storage units.
\item[$\mathcal{T},t$] Set and index of snapshots.
\item[$\mathcal{T}_k$] Set of snapshots in period $k$.
\item[$\mathcal{Z},z$] Set and index of planning sub-regions.
\item[$\cdot(b),\cdot(z)$] Set of components of the corresponding type incident to bus $b$ or sub-region $z$.
\end{IEEEdescription}
\subsection*{Subscripts and Superscripts}
\begin{IEEEdescription}[\nomlabels]
\item[0] Initial or existing value.
\item[$\mathrm{char}$] Charge.
\item[$\mathrm{dis}$] Discharge.
\item[$\mathrm{electro}$] Electrolysis.
\item[$\mathrm{fix}$] Fixed cost.
\item[$\mathrm{idle}$] Idle state.
\item[$n$] Iteration index.
\item[$\mathrm{pu}$] Per-unit.
\item[$\mathrm{sto}$] Energy storage.
\item[$\mathrm{var}$] Variable cost.
\end{IEEEdescription}
\subsection*{Parameters}
\begin{IEEEdescription}[\nomlabels]
\item[$\mathbf{A},\mathbf{b}$] Linear constraint matrix and vector of the compact problem.
\item[$a$] Availability factor.
\item[$\hat{\alpha},\hat{\beta}$] Normalized slope and intercept of AC loss-envelope segment.
\item[$C$] Cycle incidence matrix entry.
\item[$\mathbf{c}$] Cost vector of the compact problem.
\item[$c$] Annualized cost coefficient.
\item[$\chi$] Derating factor of AC line capacity.
\item[$D$] Annual demand.
\item[$d$] Electricity demand.
\item[$\Delta_t$] Snapshot duration.
\item[$\delta,\overline{\delta}$] Proximal weight and its upper bound.
\item[$\varepsilon$] Convergence tolerance.
\item[$\eta$] Efficiency.
\item[$\gamma_{\mathrm{SSSC}}$] Maximum ratio of SSSC capacity to AC line capacity.
\item[$h$] Transmission line length.
\item[$K$] Network incidence matrix entry.
\item[$\kappa$] Reactance--capacity product of AC line.
\item[$\phi$] Breakpoint of normalized AC loss segment.
\item[$\rho$] Reserve margin requirement.
\item[$\overline{U}$] Scenario-specific transmission expansion limit.
\item[$V$] Line voltage.
\item[$w$] Snapshot weight.
\end{IEEEdescription}
\subsection*{Variables}
\begin{IEEEdescription}[\nomlabels]
\item[$E$] Energy capacity.
\item[$e$] State of charge.
\item[$F$] Transmission capacity.
\item[$f$] Active power flow.
\item[$I,\overline{I}$] Line current and its upper bound.
\item[$\boldsymbol{\lambda},\boldsymbol{\mu}$] Multipliers of KVL and linear constraints.
\item[$\omega$] Progress ratio of successive capacity steps.
\item[$P$] Power capacity.
\item[$p$] Dispatch, charging power, or discharging power.
\item[$p^{\mathrm{loss}}$] Ohmic loss of AC line.
\item[$\psi$] Left-hand side of KVL constraint.
\item[$Q_{\mathrm{SSSC}}$] Three-phase SSSC reactive power capacity.
\item[$q_{\mathrm{SSSC}}$] Three-phase reactive power exchanged between the SSSC and AC line.
\item[$\tilde{q}_{\mathrm{SSSC}}$] Auxiliary SSSC control variable.
\item[$R$] Deliverable reserve.
\item[$r$] Resistance of AC line.
\item[$\overline{V}_{\mathrm{SSSC}}$] Maximum SSSC injected series voltage.
\item[$x$] Effective reactance of AC line.
\item[$x_{\mathrm{SSSC}}$] SSSC-induced series reactance on AC line.
\item[$\hat{x}$] Uncompensated reactance of AC line.
\item[$\mathbf{y}$] Vector of decision variables other than AC line capacities.
\item[$\mathbf{z}$] Vector of all decision variables.
\item[$\zeta$] Stationarity residual of the fixed-impedance solution.
\end{IEEEdescription}

\section{Introduction}
\label{sec:introduction}
\IEEEPARstart{T}{ransmission} expansion is essential for accommodating demand growth and decarbonization. National-scale U.S. studies find that transmission capacity must at least double by 2050 for cost-optimal deep decarbonization \cite{Mai2026}, yet annual additions have averaged less than 1\% of the existing network over the past 15 years \cite{FERC_Staff_Reports}, as permitting, siting, and cost allocation delay projects. Using the existing network more efficiently is therefore increasingly important. Flexible AC Transmission Systems (FACTS) can increase network transfer capability by redirecting flows from congested to underutilized lines \cite{RN1333}, and have been deployed at major interties and congested lines \cite{ENTSOE_SSSC_Technopedia,nilsson2020application}. Recent policies also promote such technologies: FERC Order No.~1920 requires transmission providers to plan over a horizon of at least 20 years under multiple scenarios and to consider alternative transmission technologies including FACTS \cite{FERC_Order1920}, and the U.S. Department of Energy has committed \$1.9 billion to various grid-enhancing technologies \cite{DOE_SPARK2026}.

Despite these operational benefits and policy support, it remains unclear how much value FACTS could provide in a nationwide grid expansion, and where and at what scale they should be deployed. These questions are intertwined with future generation and transmission investments, which shape where FACTS are most valuable and are in turn altered by FACTS deployment. Answering them therefore requires power system planning that co-optimizes FACTS with generation and transmission, consistent with the scenario-based planning that FERC Order No.~1920 now mandates \cite{FERC_Order1920}. Existing studies on power system planning with FACTS, however, lack the scale needed for nationwide planning and draw insights from stylized test or regional systems \cite{Zhang2017TPWRS, Li2018TPWRS, Franken2020IET, Hamidpour2021IJEPES, Mokhtari2021EPSR, Wu2024EPSR}. Their planning settings are not representative of real-world systems: the scenarios are illustrative rather than based on credible demand and policy projections, only transmission (and, in \cite{Li2018TPWRS, Hamidpour2021IJEPES}, conventional units) is expanded, and operation is represented by a few load levels or isolated snapshots that miss the chronological variability of demand and renewables. This lack of representativeness stems in part from the computational burden of large-scale power system planning.

In large-scale power system planning, modeling each candidate transmission investment as a discrete decision would introduce an enormous number of integer variables and render the problem intractable. A common simplification is therefore to treat line capacity expansion as a continuous decision. Under this simplification, expanding a line reduces its equivalent impedance by adding parallel circuits. This mechanism, known as \emph{impedance feedback} \cite{HAGSPIEL2014654,Lee2026TPWRS}, couples line capacities and flows in the power flow equations and makes the problem nonconvex. Existing studies typically handle it by a fixed-impedance iteration that alternates between solving the capacity expansion problem at fixed impedances and updating the impedances, which has been observed to converge \cite{HAGSPIEL2014654}. Building on this iteration, Lee and Sun \cite{Lee2026TPWRS} re-optimize line capacities in a restricted inner iteration to promote impedance consistency, and prove that a fixed point exists, but do not show that the iteration converges to it. This inner iteration is fast to solve because it decomposes by line when nodal injections are fixed, but FACTS setpoints couple all lines and remove this advantage. Critically, neither approach establishes the optimality of its solution. We will show that the fixed-impedance solution violates the Karush--Kuhn--Tucker (KKT) conditions. Thus, no existing method adequately handles the nonconvexity introduced by transmission expansion, and FACTS can add a second source of it: for a given device rating, expanding its host line reduces the series compensation the device can provide, as we show in Section~\ref{sec:pf}. We term this \emph{compensation feedback}, and refer to it together with impedance feedback as \emph{transmission feedback}, i.e., the dependence of power flows on transmission capacity decisions.

To fill these gaps, we develop a capacity expansion model of the contiguous U.S. power system that co-optimizes generation, transmission, and FACTS while handling nonconvexity. We consider SSSCs as a representative FACTS device, because, compared with continuously variable series reactors and thyristor-controlled series capacitors, they can increase and decrease the effective line reactance \cite{ENTSOE_SSSC_Technopedia}, respond quickly without generating harmonics \cite{DOE_ATT_2020}, and scale through compact modular units that are easy to install \cite{Divan2007DSSSC,EPRI_APFC_2024}. Our contributions are twofold:
\begin{enumerate}
\item We solve the nonconvex capacity expansion problem by sequential linear programming (SLP) with adaptive proximal regularization. Unlike the fixed-impedance iteration, SLP internalizes transmission feedback, i.e., how transmission expansion changes line impedances and SSSC compensation, and thus power flows. We prove that its converged solutions satisfy the KKT conditions, and show that it converges faster and reaches lower-cost expansion plans.
\item We provide the first national-scale, multi-scenario assessment of SSSC value in long-term generation and transmission expansion planning. We specifically quantify economic benefits, cost-optimal deployment patterns, and high-value deployment regions of SSSCs.
\end{enumerate}

The remainder of this paper is organized as follows. Section~\ref{sec:cem} formulates the capacity expansion problem with SSSCs, Section~\ref{sec:algorithm} presents the solution algorithms and analyzes the optimality of their converged solutions, Section~\ref{sec:results} reports the results, and Section~\ref{sec:conclusion} concludes the paper.

\section{Capacity Expansion Model with SSSCs}
\label{sec:cem}

The capacity expansion model is developed based on PyPSA \cite{Brown_PyPSA2018} and PyPSA-USA \cite{Tehranchi_PyPSAUSA2024}. Section~\ref{sec:pf} derives the SSSC-modified power flow model and the transmission feedback, and Section~\ref{sec:formulation} states the complete nonconvex planning problem.

\subsection{SSSC-Modified Power Flow Model}
\label{sec:pf}

We use the cycle-based formulation of linear power flows \cite{RN1348,Lee2026TPWRS}, as it solves faster than the conventional angle-based (B-$\theta$) and power transfer distribution factor (PTDF) formulations for large-scale capacity expansion problems. Unlike the B-$\theta$ formulation, it requires no auxiliary voltage-angle variables and fewer equality constraints; unlike the PTDF formulation, which introduces a dense PTDF matrix, it preserves the sparsity of the power flow equations. For a connected network $(\mathcal{B},\mathcal{L})$, graph theory guarantees the existence of a cycle basis $\mathcal{J}$ of cardinality $|\mathcal{L}|-|\mathcal{B}|+1$ \cite{Diestel2025}. The corresponding cycle incidence matrix $C\in\{-1,0,1\}^{|\mathcal{L}|\times|\mathcal{J}|}$ encodes the orientation of each edge relative to each cycle:
\begin{align}
C_{\ell,j}=
\begin{cases}
1  & \text{if edge } \ell \text{ is in cycle } j, \\
-1 & \text{if the reversed edge } \ell \text{ is in cycle } j, \\
0  & \text{otherwise.}
\end{cases}
\end{align}
The contiguous U.S.\ transmission network comprises three AC sub-networks, each forming an independent connected component with its own cycle basis $\mathcal{J}^{(1)}$, $\mathcal{J}^{(2)}$, and $\mathcal{J}^{(3)}$. The overall cycle basis is therefore $\mathcal{J}=\mathcal{J}^{(1)}\cup \mathcal{J}^{(2)}\cup \mathcal{J}^{(3)}$ with cardinality $|\mathcal{J}|=|\mathcal{L}|-|\mathcal{B}|+3$. Without SSSCs, Kirchhoff's voltage law (KVL) requires the reactance-weighted flows to sum to zero around each independent cycle:
\begin{equation}
\sum_{\ell\in\mathcal{L}}C_{\ell,j}\,\hat{x}^{\mathrm{pu}}_{\ell}
f^{\mathrm{pu}}_{\ell,t}=0, \quad \forall j\in\mathcal{J},\, t\in \mathcal{T}. \label{eq:kvl_base}
\end{equation}

SSSCs control the active power flow by injecting a controllable series voltage orthogonal to the line current \cite{RN1350}. Under the assumption of $V^\mathrm{pu}_\ell = 1$ and $I^\mathrm{pu}_{\ell,t} = f^\mathrm{pu}_{\ell,t}$, the effective line reactance is:
\begin{equation}
\begin{aligned}
\label{eq:sssc_x}
x^\mathrm{pu}_{\ell,t} &= \hat{x}^\mathrm{pu}_{\ell} - x^\mathrm{pu}_{\text{SSSC},\ell,t} = \hat{x}^\mathrm{pu}_{\ell} - \frac{q^\mathrm{pu}_{\text{SSSC},\ell,t}}{{I_{\ell,t}^\mathrm{pu}}^2} \\
&=\hat{x}^\mathrm{pu}_{\ell} - \frac{q^\mathrm{pu}_{\text{SSSC},\ell,t}}{{f_{\ell,t}^\mathrm{pu}}^2}, \quad
\forall \ell \in \mathcal{L}, \, t \in \mathcal{T}.
\end{aligned}
\end{equation}
Because the SSSC is connected in series with the transmission line, its injected-voltage capability is tied to the line's maximum operating current:
\begin{equation}
\label{eq:v_sssc}
\overline V^\mathrm{pu}_{\text{SSSC},\ell} = \frac{Q^\mathrm{pu}_{\text{SSSC},\ell}}{\sqrt{3}\,\overline{I}^\mathrm{pu}_\ell} = \frac{Q^\mathrm{pu}_{\text{SSSC},\ell}}{\sqrt{3}\,F^\mathrm{pu}_\ell}, \quad \forall \ell \in \mathcal{L}.
\end{equation}
Under varying line flow conditions, the SSSC reactive power injection is bounded by the product of the maximum injected voltage and the line current:
\begin{equation}
\begin{aligned}
\label{eq:q_limit}
|q^\mathrm{pu}_{\text{SSSC},\ell,t}| \le |\sqrt{3}\,\overline V^\mathrm{pu}_{\text{SSSC},\ell}\, I^\mathrm{pu}_{\ell,t}| = Q^\mathrm{pu}_{\text{SSSC},\ell}\, \frac{|f^\mathrm{pu}_{\ell,t}|}{F^\mathrm{pu}_\ell}, \\ \forall \ell \in \mathcal{L},  \, t \in \mathcal{T}.
\end{aligned}
\end{equation}
Substituting \eqref{eq:sssc_x} into \eqref{eq:kvl_base} and introducing the auxiliary variable $\tilde{q}^\mathrm{pu}_{\mathrm{SSSC},\ell,t} := q^\mathrm{pu}_{\mathrm{SSSC},\ell,t}\,F^\mathrm{pu}_\ell/f^\mathrm{pu}_{\ell,t}$ yields the SSSC-modified KVL
\begin{align}
\sum_{\ell\in\mathcal{L}} C_{\ell,j}\, \left( \hat{x}^\mathrm{pu}_{\ell}\, f^\mathrm{pu}_{\ell,t} -  \frac{\tilde{q}^\mathrm{pu}_{\mathrm{SSSC},\ell,t}}{F^\mathrm{pu}_{\ell}} \right) = 0, \quad \forall j \in \mathcal{J},\, t \in \mathcal{T},
\label{eq:kvl_sssc}
\end{align}
and \eqref{eq:q_limit} reduces to the box constraint $|\tilde{q}^\mathrm{pu}_{\mathrm{SSSC},\ell,t}| \leq Q^\mathrm{pu}_{\mathrm{SSSC},\ell}$. For a fixed line capacity, \eqref{eq:kvl_sssc} is linear in the flows and SSSC controls.

Line capacity, however, is a decision variable, and it affects~\eqref{eq:kvl_sssc} through two channels. First, line impedance decreases as transmission expands, a mechanism known as \emph{impedance feedback} \cite{HAGSPIEL2014654,Lee2026TPWRS}:
\begin{align}
r^\mathrm{pu}_\ell = r_{\ell}^{\mathrm{pu},0}\,\frac{F_{\ell}^0}{F_\ell}, \qquad
\hat{x}^\mathrm{pu}_\ell = \hat{x}_{\ell}^{\mathrm{pu},0}\,\frac{F_{\ell}^0}{F_\ell}, \qquad \forall \ell \in \mathcal{L}.
\label{eq:impedance_update}
\end{align}
Second, by~\eqref{eq:v_sssc}, a given SSSC rating supports a lower injected voltage on a line of larger capacity. Equivalently, the SSSC term $\tilde{q}^\mathrm{pu}_{\mathrm{SSSC},\ell,t}/F^\mathrm{pu}_\ell$ in~\eqref{eq:kvl_sssc} is bounded in magnitude by the compensation ratio $Q_{\mathrm{SSSC},\ell}/F_\ell$, so expanding a line at a given SSSC rating reduces the series compensation the SSSC can provide. We term this \emph{compensation feedback}, and refer to the two channels jointly as \emph{transmission feedback}, which reduces to impedance feedback without SSSCs. Substituting \eqref{eq:impedance_update} into \eqref{eq:kvl_sssc} gives the nonlinear KVL constraint \eqref{eq:kvl} of the planning problem below, which makes the problem nonconvex.

\subsection{Capacity Expansion Problem Formulation}
\label{sec:formulation}

The capacity expansion problem with SSSC investment and operation is formulated as follows:
\allowdisplaybreaks
\begin{subequations}
\label{eq:cem}
\begin{align}
\min \,\,
& \sum_{g\in\mathcal{G}} \Bigl(c^\mathrm{fix}_g P_g + c^\mathrm{var}_{g}\sum_{t\in \mathcal{T}}w_t p_{g,t}\Bigr)
+ \sum_{i\in\mathcal{I}} c_{i}F_{i} \nonumber \\
&+ \sum_{\ell\in\mathcal{L}} \bigl(c_{\ell}F_{\ell} + c_\mathrm{SSSC}\, Q_{\mathrm{SSSC},\ell}\bigr) \nonumber \\
&+ \sum_{s\in\mathcal{S}} \bigl(c^\mathrm{char}_{s}P^\mathrm{char}_{s} + c^\mathrm{dis}_{s}P^\mathrm{dis}_{s} + c^\mathrm{sto}_{s}E_s\bigr) \label{eq:obj_fun}\\
\text{s.t.}\,\,
&\sum_{g\in\mathcal{G}(b)} p_{g,t} + \sum_{\ell\in\mathcal{L}(b)}\!\left(K^\mathrm{AC}_{\ell,b} f_{\ell,t}
- \tfrac{1}{2}p^{\mathrm{loss}}_{\ell,t}\right) + \sum_{i\in\mathcal{I}(b)} K^\mathrm{DC}_{i,b}f_{i,t} \nonumber \\
&+ \sum_{s\in\mathcal{S}(b)}(p^\mathrm{dis}_{s,t}-p^\mathrm{char}_{s,t})
= d_{b,t}, \quad \forall b\in\mathcal{B},\, t\in \mathcal{T}, \label{eq:nodal_balance}\\
&\sum_{t\in \mathcal{T}}w_t\sum_{g\in\mathcal{G}_{\mathrm{electro}}(z)}-p_{g,t}
= D_{\mathrm{electro},z}, \quad \forall z\in\mathcal{Z}, \label{eq:h2_balance}\\
&P_g,F_\ell,F_i,P^\mathrm{char}_s,P^\mathrm{dis}_s,E_s,Q_{\mathrm{SSSC},\ell} \ge 0, \nonumber \\
&\forall g\in\mathcal{G},\,\ell\in\mathcal{L},\,i\in\mathcal{I},\,s\in\mathcal{S}, \label{eq:cap_nn}\\
&P_{g} \le \overline{P}_{g},
\quad \forall g \in \mathcal{G}, \label{eq:gen_limit}\\
&F_\ell^0 \le F_\ell \le \overline{F}_\ell,
\quad \forall \ell\in\mathcal{L}, \label{eq:ac_limit}\\
&F_i^0 \le F_i \le \overline{F}_i,
\quad \forall i\in\mathcal{I}, \label{eq:dc_limit}\\
&\sum_{\ell\in\mathcal{L}}h_\ell(F_\ell-F_\ell^0)
+\sum_{i\in\mathcal{I}}\tfrac{1}{2}h_i(F_i-F_i^0)
\le \overline{U}, \label{eq:trans_limit}\\
&F_i = F_{i'},
\quad \forall i\in\mathcal{I}, \label{eq:dc_symmetry}\\
&0 \le p_{g,t} \le a_{g,t}P_g, \quad \forall g\notin\mathcal{G}_{\mathrm{electro}},\, t\in \mathcal{T},\label{eq:dispatch_g}\\
&-P_g \le p_{g,t} \le 0,
\quad \forall g\in\mathcal{G}_{\mathrm{electro}},\, t\in \mathcal{T}, \label{eq:dispatch_e}\\
&0 \le p^\mathrm{char}_{s,t} \le P^\mathrm{char}_s,
\quad \forall s\in\mathcal{S},\, t\in \mathcal{T}, \label{eq:dispatch_s1}\\
&0 \le p^\mathrm{dis}_{s,t} \le P^\mathrm{dis}_s,
\quad \forall s\in\mathcal{S},\, t\in \mathcal{T}, \label{eq:dispatch_s2}\\
&0 \le f_{i,t} \le F_i,
\quad \forall i\in\mathcal{I},\, t\in \mathcal{T}, \label{eq:dispatch_i}\\
&e_{s,k,t} = {\eta^\mathrm{idle}_{s}}^{\Delta_t}e_{s,k,t-1}
+\Delta_t\eta^\mathrm{char}_{s}p^\mathrm{char}_{s,k,t} -\frac{\Delta_t}{\eta^\mathrm{dis}_{s}}p^\mathrm{dis}_{s,k,t}, \nonumber \\
&\hspace{1em}\forall s\in\mathcal{S},\, k\in\mathcal{K},\, t\in\mathcal{T}_k, \label{eq:soc_dynamics}\\
&0 \le e_{s,k,t} \le E_s,
\quad \forall s\in\mathcal{S},\, k\in\mathcal{K},\, t\in\mathcal{T}_k, \label{eq:soc_limit}\\
&e_{s,k,0}=e_{s,k,|\mathcal{T}_k|},
\quad \forall s\in\mathcal{S},\, k\in\mathcal{K}, \label{eq:cyclic_soc}\\
&\sum_{\ell\in\mathcal{L}} C_{\ell,j}\,
\frac{\kappa_\ell\, f^\mathrm{pu}_{\ell,t} - \tilde{q}^\mathrm{pu}_{\mathrm{SSSC},\ell,t}}{F^\mathrm{pu}_\ell} = 0,
\quad \forall j\in\mathcal{J},\, t\in \mathcal{T}, \label{eq:kvl}\\
&p^{\mathrm{loss}}_{\ell,t} \ge r^{\mathrm{pu},0}_\ell F^{\mathrm{pu},0}_\ell\bigl(\hat{\alpha}_{m}f_{\ell,t}+\hat{\beta}_{m}F_\ell\bigr), \nonumber\\
&\hspace{1em}\forall \ell\in\mathcal{L},\, t\in \mathcal{T},\, m\in\mathcal{M}, \label{eq:loss_envelope}\\
&|f_{\ell,t}|+p^{\mathrm{loss}}_{\ell,t}\le \chi F_\ell,
\quad \forall \ell\in\mathcal{L},\, t\in \mathcal{T}, \label{eq:thermal_limit}\\
&|\tilde{q}^\mathrm{pu}_{\mathrm{SSSC},\ell,t}| \leq Q^\mathrm{pu}_{\mathrm{SSSC},\ell}, \quad \forall \ell \in \mathcal{L},\ t \in \mathcal{T}, \label{eq:q_bound}\\
&Q_{\mathrm{SSSC},\ell} \leq \gamma_{\mathrm{SSSC}}\,F_\ell, \quad \forall \ell \in \mathcal{L}, \label{eq:sssc_rating}\\
&\sum_{b\in\mathcal{B}(z)}\biggl[
\sum_{g\in\mathcal{G}(b)}a_{g,t}P_g
+\sum_{s\in\mathcal{S}(b)}R_{s,t} \nonumber\\
&\qquad+\sum_{\ell\in\mathcal{L}(b)}
\left(K^\mathrm{AC}_{\ell,b}f_{\ell,t}-\tfrac{1}{2}p^{\mathrm{loss}}_{\ell,t}\right)
+\sum_{i\in\mathcal{I}(b)}K^\mathrm{DC}_{i,b}f_{i,t}\biggr] \nonumber\\
&\ge (1+\rho)\sum_{b\in\mathcal{B}(z)}d_{b,t}, \quad \forall z\in\mathcal{Z},\, t\in \mathcal{T}, \label{eq:reserve_margin}\\
&0\le R_{s,t}\le P^\mathrm{dis}_s,\quad
R_{s,t}\le \frac{e_{s,t}\eta^\mathrm{dis}_s}{\Delta_t}, \quad \forall s\in\mathcal{S},\, t\in \mathcal{T}. \label{eq:erm_s}
\end{align}
\end{subequations}

Problem~\eqref{eq:cem} seeks optimal investment decisions including generation capacity $P_g$, AC and DC transmission capacity $F_\ell$ and $F_i$, storage power and energy capacity $P^\mathrm{char}_s$, $P^\mathrm{dis}_s$, $E_s$, and SSSC capacity $Q_{\mathrm{SSSC},\ell}$, jointly with operational decisions including generation and electrolysis dispatch $p_{g,t}$; storage charging, discharging, and state of charge $p^\mathrm{char}_{s,t}$, $p^\mathrm{dis}_{s,t}$, $e_{s,k,t}$; AC and DC power flows $f_{\ell,t}$ and $f_{i,t}$ with associated ohmic losses; and SSSC controls $\tilde{q}^\mathrm{pu}_{\mathrm{SSSC},\ell,t}$.

Equation~\eqref{eq:obj_fun} is the optimization objective, which minimizes total annualized system cost from generation and electrolysis capacity, generation dispatch, transmission capacity, storage power and energy capacity, and SSSC capacity. Equation~\eqref{eq:nodal_balance} enforces nodal energy balance. The AC incidence matrix $K^\text{AC}_{\ell,b} \in \{-1,0,+1\}$ equals $-1$ if line $\ell$ originates at bus $b$ and $+1$ if it terminates there. The DC incidence matrix $K^\text{DC}_{i,b} \in \{-1,0,\eta_i\}$ equals $-1$ if link $i$ originates at bus $b$, and $\eta_i$ if it terminates there. Equation~\eqref{eq:h2_balance} imposes the annual flexible electrolysis load separately for each planning sub-region $z\in\mathcal{Z}$ \cite{NREL_ReEDS}.

Equations~\eqref{eq:cap_nn}--\eqref{eq:dc_limit} impose lower and upper capacity limits for generation, storage, transmission, and SSSCs. Equation~\eqref{eq:trans_limit} limits the total transmission expansion volume in each scenario. Each DC line is represented by a pair of unidirectional links whose capacities are equalized by~\eqref{eq:dc_symmetry}, so each link is weighted by one half in~\eqref{eq:trans_limit} to count the line once. For transmission lines, $F$ includes both existing and expanded capacity.

Equations~\eqref{eq:dispatch_g}--\eqref{eq:dispatch_i} bound dispatch from generators, electrolyzers, storage units, and DC lines. Equations~\eqref{eq:soc_dynamics}--\eqref{eq:cyclic_soc} enforce storage energy dynamics, storage energy capacity limits, and cyclic consistency within each period. 

Equation~\eqref{eq:kvl} is the SSSC-modified KVL derived in Section~\ref{sec:pf}, where $\kappa_\ell := \hat{x}^{\mathrm{pu},0}_\ell F^{\mathrm{pu},0}_\ell$ is the capacity-invariant reactance--capacity product. It is the only nonlinear constraint in~\eqref{eq:cem} and makes Problem~\eqref{eq:cem} nonconvex. Both channels of transmission feedback enter through the common factor $1/F^\mathrm{pu}_\ell$. The first term, $\kappa_\ell f^\mathrm{pu}_{\ell,t}/F^\mathrm{pu}_\ell$, represents impedance feedback: expanding a line lowers its reactance and attracts flow to it. The second term, $\tilde{q}^\mathrm{pu}_{\mathrm{SSSC},\ell,t}/F^\mathrm{pu}_\ell$, represents compensation feedback: for a given SSSC rating, expanding a line weakens the series compensation its SSSC can provide.

Equation~\eqref{eq:loss_envelope} approximates the quadratic AC ohmic loss $r^\mathrm{pu}_\ell f^2$ over the admissible flow range $f\in[-\chi F_\ell,\chi F_\ell]$ by $|\mathcal{M}|$ least-squares segments. Its normalized coefficients are fitted on the normalized flow range $[-\chi,\chi]$ partitioned by breakpoints $-\chi=\phi_1<\dots<\phi_{|\mathcal{M}|+1}=\chi$:
\begin{equation}
\hat{\alpha}_{m}, \hat{\beta}_{m} = \operatorname*{argmin}_{\alpha,\beta} \int_{\phi_{m}}^{\phi_{m+1}} \left( \phi^2 - (\alpha \phi + \beta) \right)^2 d\phi, \label{eq:ab}
\end{equation}
with $\beta=0$ imposed on the innermost segment so that zero flow incurs zero loss, where the derating factor $\chi$, set to 0.7, approximates N$-1$ security constraints \cite{gazafroudi2022topology}. Because the resistance in~\eqref{eq:impedance_update} and the segment breakpoints both scale with $F_\ell$, \eqref{eq:loss_envelope} is linear in $(f_{\ell,t},F_\ell)$ and keeps the same form across the iterations of Section~\ref{sec:algorithm}. Equation~\eqref{eq:thermal_limit} enforces AC thermal limits. Equation~\eqref{eq:q_bound} bounds SSSC control by the installed SSSC capacity. Equation~\eqref{eq:sssc_rating} limits the SSSC rating on each line to a fraction $\gamma_{\mathrm{SSSC}}=0.3$ of the line capacity \cite{Singh_EPSR_2021}.

Equations~\eqref{eq:reserve_margin}--\eqref{eq:erm_s} enforce the reserve margin $\rho$ separately for each planning sub-region. Generation and storage contribute their maximum deliverable capacity, while transmission contributes actual net power inflows.

\section{Solution Algorithm}
\label{sec:algorithm}

Problem~\eqref{eq:cem} is solved as a sequence of convex subproblems in which~\eqref{eq:kvl} is approximated around the current iterate. Section~\ref{sec:fi} reviews the conventional fixed-impedance iteration, and Section~\ref{sec:slp} develops sequential linear programming with adaptive proximal regularization, which improves on the former in terms of optimality (Section~\ref{sec:kkt}).

Let $\mathbf{z}$ collect all decision variables of~\eqref{eq:cem}, and let $\mathbf{F}=(F_\ell)_{\ell\in\mathcal{L}}$ denote the AC line capacities and $\mathbf{y}$ the remaining variables, so that $\mathbf{z}=(\mathbf{F},\mathbf{y})$. Problem~\eqref{eq:cem} can be written compactly as
\begin{equation}
\begin{aligned}
\min_{\mathbf{z}}\ & \mathbf{c}^\top\mathbf{z} \\
\text{s.t.}\ & \mathbf{A}\mathbf{z}\le\mathbf{b} && :\ \boldsymbol{\mu}, \\
& \boldsymbol{\psi}(\mathbf{z})=\mathbf{0} && :\ \boldsymbol{\lambda},
\end{aligned}
\label{eq:compact}
\end{equation}
where $\mathbf{A}\mathbf{z}\le\mathbf{b}$ gathers all linear constraints and $\boldsymbol{\psi}$ stacks the left-hand sides $\psi_{j,t}(\mathbf{z})$ of~\eqref{eq:kvl}, with
\begin{equation}
\frac{\partial \psi_{j,t}}{\partial F^\mathrm{pu}_\ell}
= -\,C_{\ell,j}\,\frac{\kappa_\ell f^\mathrm{pu}_{\ell,t}-\tilde{q}^\mathrm{pu}_{\mathrm{SSSC},\ell,t}}{(F^\mathrm{pu}_\ell)^2}. \label{eq:dpsi}
\end{equation}

\subsection{Fixed-Impedance Iteration}
\label{sec:fi}

The conventional approach to impedance feedback is a fixed-impedance (FI) iteration \cite{HAGSPIEL2014654}. At iteration $n$, the line capacities in~\eqref{eq:kvl} are frozen at the previous iterate $\mathbf{F}^n$:
\begin{equation}
\sum_{\ell\in\mathcal{L}} C_{\ell,j}\,
\frac{\kappa_\ell\, f^\mathrm{pu}_{\ell,t} - \tilde{q}^\mathrm{pu}_{\mathrm{SSSC},\ell,t}}{F^{\mathrm{pu},n}_\ell} = 0,
\quad \forall j\in\mathcal{J},\, t\in \mathcal{T}. \label{eq:kvl_fi}
\end{equation}
With SSSCs, freezing $\mathbf{F}^n$ also freezes the compensation each unit of SSSC rating provides. The resulting LP is solved for $\mathbf{z}^{n+1}$, and the procedure repeats until the system cost and capacities stop changing, as summarized in Algorithm~\ref{alg:fi}. The iteration terminates when both the relative change in system cost and the relative capacity step $\|\mathbf{F}^{n+1}-\mathbf{F}^n\|_1/\|\mathbf{F}^0\|_1$ fall below their tolerances. The FI iteration enforces transmission feedback only between iterations: each subproblem treats line impedances and SSSC compensation per unit rating as exogenous and ignores how its own capacity decisions change them.

\begin{algorithm}[t]
\caption{Fixed-Impedance Iteration}
\label{alg:fi}
\begin{algorithmic}[1]
\Require Existing capacities $\mathbf{F}^0$; tol. $\varepsilon_{\mathrm{obj}}=10^{-5}$, $\varepsilon_F=10^{-2}$
\Ensure Investment and operation decisions
\State $n\gets 0$
\Repeat
    \State Build~\eqref{eq:kvl_fi} at $\mathbf{F}^n$
    \State Solve~\eqref{eq:cem} with~\eqref{eq:kvl} replaced by~\eqref{eq:kvl_fi} to obtain $\mathbf{z}^{n+1}$
    \State $n\gets n+1$
\Until{cost change $\le\varepsilon_{\mathrm{obj}}$ \textbf{and} $\|\mathbf{F}^{n}-\mathbf{F}^{n-1}\|_1\le\varepsilon_F\|\mathbf{F}^0\|_1$}
\State Fix $\mathbf{F}\gets\mathbf{F}^n$ and solve~\eqref{eq:cem} to obtain an impedance-consistent solution
\State \Return the solution of the final solve
\end{algorithmic}
\end{algorithm}

\subsection{Sequential Linear Programming with Proximal Regularization}
\label{sec:slp}

\begin{algorithm}[t]
\caption{SLP with Adaptive Proximal Regularization}
\label{alg:slp}
\begin{algorithmic}[1]
\Require Existing capacities $\mathbf{F}^0$; $\delta^1=0.5$, $\overline{\delta}=4$; tol. $\varepsilon_{\mathrm{obj}}=10^{-5}$, $\varepsilon_F=10^{-2}$
\Ensure Investment and operation decisions
\State Solve~\eqref{eq:cem} with~\eqref{eq:kvl} replaced by~\eqref{eq:kvl_fi} at $\mathbf{F}^0$ to obtain $\mathbf{z}^1$; $n\gets 1$
\Repeat
    \State Build~\eqref{eq:kvl_slp} at $\mathbf{z}^n$ and the proximal term at $\mathbf{F}^n$ with weight $\delta^n$
    \State Solve~\eqref{eq:slp} to obtain $\mathbf{z}^{n+1}$
    \If{$n\ge 2$ \textbf{and} $\omega^n<0.75$}
        \State $\delta^{n+1}\gets\min(2\delta^n,\overline{\delta})$
    \Else
        \State $\delta^{n+1}\gets\delta^n$
    \EndIf
    \State $n\gets n+1$
\Until{cost change $\le\varepsilon_{\mathrm{obj}}$ \textbf{and} $\|\mathbf{F}^{n}-\mathbf{F}^{n-1}\|_1\le\varepsilon_F\|\mathbf{F}^0\|_1$}
\State Fix $\mathbf{F}\gets\mathbf{F}^n$ and solve~\eqref{eq:cem} to obtain an impedance-consistent solution
\State \Return the solution of the final solve
\end{algorithmic}
\end{algorithm}

SLP \cite{palacios1982nonlinear} instead replaces~\eqref{eq:kvl} with its first-order Taylor expansion at $\mathbf{z}^n=(\mathbf{F}^n,\mathbf{y}^n)$:
\begin{equation}
\begin{aligned}
\sum_{\ell\in\mathcal{L}} C_{\ell,j}\Biggl[&
\frac{\kappa_\ell f^\mathrm{pu}_{\ell,t} - \tilde{q}^\mathrm{pu}_{\mathrm{SSSC},\ell,t}}{F^{\mathrm{pu},n}_\ell} \\
&-\frac{\kappa_\ell f^{\mathrm{pu},n}_{\ell,t} - \tilde{q}^{\mathrm{pu},n}_{\mathrm{SSSC},\ell,t}}{(F^{\mathrm{pu},n}_\ell)^2}
\bigl(F^\mathrm{pu}_\ell-F^{\mathrm{pu},n}_\ell\bigr)\Biggr] = 0, \\
&\forall j\in\mathcal{J},\, t\in \mathcal{T}.
\end{aligned}
\label{eq:kvl_slp}
\end{equation}
Because~\eqref{eq:kvl} is linear in $(f,\tilde{q})$ for fixed $F$, the first term of~\eqref{eq:kvl_slp} coincides with the FI constraint~\eqref{eq:kvl_fi}; the two differ only in the last term, which is the capacity sensitivity~\eqref{eq:dpsi} evaluated at $\mathbf{z}^n$. This term internalizes transmission feedback: its $\kappa_\ell f^{\mathrm{pu},n}_{\ell,t}$ part anticipates impedance feedback, and its $\tilde{q}^{\mathrm{pu},n}_{\mathrm{SSSC},\ell,t}$ part anticipates compensation feedback.

The linearization is accurate only near $\mathbf{F}^n$. We therefore add a quadratic proximal term on the AC line capacities, and solve at each iteration the convex quadratic program
\begin{equation}
\begin{aligned}
\min_{\mathbf{z}}\ & \mathbf{c}^\top\mathbf{z}
+ \delta^n \sum_{\ell\in\mathcal{L}} c_\ell\,\frac{(F_\ell-F^n_\ell)^2}{F^n_\ell} \\
\text{s.t.}\ & \mathbf{A}\mathbf{z}\le\mathbf{b}, \quad \text{\eqref{eq:kvl_slp}},
\end{aligned}
\label{eq:slp}
\end{equation}
where $c_\ell$ is the annualized capacity cost of line $\ell$ in~\eqref{eq:obj_fun}. The normalization by $F^n_\ell$ makes the weight $\delta^n$ dimensionless: moving the capacity of a line by its own size costs $\delta^n$ times the annualized capital cost of that line. Moreover, the gradient of the proximal term vanishes at $\mathbf{F}=\mathbf{F}^n$, so the proximal term does not alter the KKT conditions at convergence, and any converged solution of SLP is a KKT point for any bounded sequence $\{\delta^n\}$ (Proposition~\ref{prop:slp}).

The weight $\delta^n$ is adapted according to whether the iterates progress steadily or oscillate, which is detected by the progress ratio
\begin{equation}
\omega^n = \frac{\sum_{\ell} c_\ell\,\bigl|F^{n+1}_\ell-F^{n-1}_\ell\bigr|}
{\sum_{\ell} c_\ell\bigl(|F^{n+1}_\ell-F^{n}_\ell|+|F^{n}_\ell-F^{n-1}_\ell|\bigr)} \in [0,1]. \label{eq:progress}
\end{equation}
It divides the cost-weighted net transmission capacity change over the last two iterations by the sum of the changes in the two individual steps. It equals one if every transmission capacity moves in the same direction in both steps and zero if every transmission capacity at iteration $n+1$ returns to its value at iteration $n-1$. When $\omega^n<0.75$ (manually tuned threshold), the iteration is deemed oscillating and $\delta^n$ is doubled, up to $\overline{\delta}$. Algorithm~\ref{alg:slp} summarizes the procedure. The first iteration is a fixed-impedance step from the existing capacities $\mathbf{F}^0$, because no linearization point is available yet. The termination criteria and the final solve are the same as in Algorithm~\ref{alg:fi}, with the system cost evaluated without the proximal term.

\subsection{Optimality of the Converged Solutions}
\label{sec:kkt}

A point $\mathbf{z}$ is a KKT point of~\eqref{eq:compact} if there exist dual variables $\boldsymbol{\lambda}$ and $\boldsymbol{\mu}$ as indicated in~\eqref{eq:compact} such that
\begin{subequations}
\label{eq:kkt}
\begin{align}
&\mathbf{c}+\nabla\boldsymbol{\psi}(\mathbf{z})^\top\boldsymbol{\lambda}+\mathbf{A}^\top\boldsymbol{\mu}=\mathbf{0}, \label{eq:kkt_stat}\\
&\boldsymbol{\psi}(\mathbf{z})=\mathbf{0},\quad \mathbf{A}\mathbf{z}\le\mathbf{b}, \label{eq:kkt_feas}\\
&\boldsymbol{\mu}\ge\mathbf{0}, \label{eq:kkt_dual}\\
&\boldsymbol{\mu}^\top(\mathbf{A}\mathbf{z}-\mathbf{b})=0. \label{eq:kkt_comp}
\end{align}
\end{subequations}
Although~\eqref{eq:cem} is nonconvex, the KKT conditions are necessary for a local minimum under standard constraint qualifications, and thus serve as the yardstick for comparing the two schemes.

Both algorithms solve a convex subproblem at each iteration. Let $\boldsymbol{\lambda}^n$ and $\boldsymbol{\mu}^n$ denote the corresponding dual variables of the subproblem at iteration $n$, with $\boldsymbol{\lambda}^n$ associated with~\eqref{eq:kvl_fi} or~\eqref{eq:kvl_slp}. We consider the case in which the iterates and these dual variables converge, $\mathbf{z}^n\to\mathbf{z}^\star$ and $(\boldsymbol{\lambda}^n,\boldsymbol{\mu}^n)\to(\boldsymbol{\lambda}^\star,\boldsymbol{\mu}^\star)$, and ask whether the converged solution $\mathbf{z}^\star$ is a KKT point of~\eqref{eq:compact}.

\begin{proposition}[Fixed-impedance iteration]
\label{prop:fi}
The converged solution of Algorithm~\ref{alg:fi} satisfies all KKT conditions~\eqref{eq:kkt} except stationarity with respect to $\mathbf{F}$, which is violated by
\begin{equation}
\zeta_\ell=-\sum_{t\in \mathcal{T}}\sum_{j\in\mathcal{J}}\lambda^\star_{j,t}\,C_{\ell,j}\,
\frac{\kappa_\ell f^{\mathrm{pu},\star}_{\ell,t}-\tilde{q}^{\mathrm{pu},\star}_{\mathrm{SSSC},\ell,t}}{(F^{\mathrm{pu},\star}_\ell)^2}, \quad \forall \ell\in\mathcal{L}. \label{eq:zeta}
\end{equation}
\end{proposition}

\begin{IEEEproof}
The FI constraint~\eqref{eq:kvl_fi} is $\boldsymbol{\psi}(\mathbf{F}^n,\mathbf{y})=\mathbf{0}$: it has the same value and the same gradient with respect to $\mathbf{y}$ as $\boldsymbol{\psi}$, but a zero gradient with respect to $\mathbf{F}$. The KKT conditions of the FI subproblem at convergence yield~\eqref{eq:kkt_feas}--\eqref{eq:kkt_comp} and~\eqref{eq:kkt_stat} with respect to $\mathbf{y}$. With respect to $\mathbf{F}$, the term $\nabla_{\mathbf{F}}\boldsymbol{\psi}(\mathbf{z}^\star)^\top\boldsymbol{\lambda}^\star$ is missing, which equals $\boldsymbol{\zeta}$ by~\eqref{eq:dpsi}.
\end{IEEEproof}

\begin{proposition}[SLP with quadratic proximal regularization]
\label{prop:slp}
The converged solution of Algorithm~\ref{alg:slp} satisfies all KKT conditions~\eqref{eq:kkt}.
\end{proposition}

\begin{IEEEproof}
Unlike~\eqref{eq:kvl_fi}, the linearized constraint~\eqref{eq:kvl_slp} matches both the value and the full gradient of $\boldsymbol{\psi}$ at $\mathbf{z}^n$. The proximal term adds $2\delta^n c_\ell(F^{n+1}_\ell-F^n_\ell)/F^n_\ell$ to stationarity with respect to $F_\ell$, which vanishes at convergence. The KKT conditions of~\eqref{eq:slp} at convergence, as in the proof of Proposition~\ref{prop:fi}, therefore recover~\eqref{eq:kkt} in full.
\end{IEEEproof}

Propositions~\ref{prop:fi} and~\ref{prop:slp} identify the difference between the two algorithms. Whenever it converges, SLP with quadratic proximal regularization reaches a KKT point of~\eqref{eq:cem}. The FI iteration instead reaches a point at which no line expansion is profitable at frozen impedances. This point violates stationarity of~\eqref{eq:cem} by $\boldsymbol{\zeta}$, and is therefore not a stationary point of~\eqref{eq:cem}. The transmission correction of~\cite{Lee2026TPWRS} shares this limitation, because it also freezes the impedances within each re-optimization and, in addition, keeps all non-transmission decisions fixed. The residual $\zeta_\ell$ in~\eqref{eq:zeta} is the marginal system cost of transmission feedback from expanding line $\ell$, weighted by the KVL shadow prices $\lambda^\star_{j,t}$. The two terms of $\zeta_\ell$ correspond to impedance feedback and compensation feedback in~\eqref{eq:kvl}, whose marginal effects the FI iteration ignores and SLP internalizes.

\section{Scenarios and Results}
\label{sec:results}
\subsection{System Configuration and Scenario Design}
\label{sec:setup}

The transmission network topology and voltage levels for the three interconnections of the contiguous U.S. are sourced from the substation-level Homeland Infrastructure Foundation-Level Data (HIFLD) \cite{HIFLD}, which covers transmission lines at 69~kV and above. The typical capacity and per-unit-length impedance of each line are assigned according to its voltage level based on the TAMU test system \cite{xu2020us}. The substation-level HIFLD network comprises 52,476 buses. We first merge degree-1 buses into their adjacent buses and eliminate degree-2 buses that have no injection and connect two series lines of the same voltage level \cite{Grudzien2018}. Because flows on radial branches cannot be redirected, and two series lines without an intermediate injection carry identical flows and are exactly represented by one equivalent line, this step reduces the network to 27,399 buses without affecting SSSC valuation. The network is further reduced to 5,000 buses by aggregation based on effective electrical distance \cite{Spielman2011} and physical distance between buses. Lines connecting the same pair of buses after aggregation are merged into one line whose capacity is the sum of their capacities and whose impedance is their parallel combination. Hereafter, a line refers to a branch of the reduced network, which may represent several parallel circuits or series segments of the physical network. SSSCs are modeled on the merged lines. This representation is exact for parallel circuits with identical reactance--capacity products $\kappa_\ell$ when the SSSC capacity is allocated in proportion to circuit capacities, and it is a good approximation otherwise. Our validation shows that the system cost, cost savings from SSSCs, and optimal SSSC capacity obtained with the 5,000-bus network differ from those obtained with the 27,399-bus network by approximately 0.4\%, 3.3\%, and 0.8\%, respectively. These deviations are acceptable for the purposes of this study.

The model is initialized using the 2026 fleet of generation and storage units \cite{Selvans2026_PUDL}. Techno-economic parameters are sourced from the 2025 NLR Annual Technology Baseline \cite{NLR_ATB_2025} and the 2024 Annual Decarbonization Perspective \cite{EvolvedEnergy_ADP2024}. Route-specific transmission expansion costs per MW-mile are obtained from ReEDS \cite{NREL_ReEDS}. The cost of SSSCs is estimated based on project data submitted by Smart Wires, Inc.\ to CAISO \cite{smartwires2019}. System operation is represented by 10 representative days selected from 15 weather years of demand and renewable data \cite{NREL_ReEDS}. These include three extreme days with the maximum load, the lowest wind availability, and the lowest solar availability, and seven typical days. Each representative day is modeled at 2-hour resolution, and snapshot weights $w$ are set so that they sum to 8760. Interday flexibility is implicitly captured by flexible electrolyzer operation \cite{ai_delineating}.

We construct a baseline scenario for 2050 and optimize with and without SSSCs, so that their value is measured by the difference between the two solutions. In the baseline, annual exogenous electricity demand is approximately 6400~TWh \cite{NREL_ReEDS}. Decarbonization follows the current federal and state-level policies, namely the Clean Air Act, Technology Capacity Targets, Renewable Portfolio Standards, and Clean Energy Standards \cite{NREL_ReEDS}, along with the mandatory retirement of oil- and coal-fired power plants by 2050. A 95\% reduction in power-sector greenhouse gas emissions relative to 2005 is required. The total annual flexible electrolysis load is 476~TWh \cite{EvolvedEnergy_ADP2024}. Total transmission expansion is limited to 20\% of the current network volume, which corresponds to continuing the average expansion rate of the past 15 years \cite{FERC_Staff_Reports}. The cost of SSSCs is 60~\$/kVAr \cite{smartwires2019}. The alternative transmission expansion, demand, decarbonization, and SSSC cost scenarios are introduced with their corresponding results below.

\subsection{Algorithm Performance}
\label{sec:res_alg}

We first compare the FI iteration (Algorithm~\ref{alg:fi}) and SLP with adaptive proximal regularization (Algorithm~\ref{alg:slp}) on a demonstration case with reduced temporal fidelity. Two cost measures are tracked at each iteration $n$. The impedance-inconsistent cost is the objective value reported by the subproblem that produces $\mathbf{F}^{n}$. Because the KVL constraints of this subproblem are built at the previous iterate, either frozen in~\eqref{eq:kvl_fi} or linearized in~\eqref{eq:kvl_slp}, the power flows underlying this cost are not consistent with the impedances implied by $\mathbf{F}^{n}$ itself. The impedance-consistent cost is obtained by fixing the AC line capacities at $\mathbf{F}^{n}$, updating the line impedances according to~\eqref{eq:impedance_update}, and re-optimizing all remaining decisions.

\begin{figure}[t]
\centering
\includegraphics[width=\columnwidth]{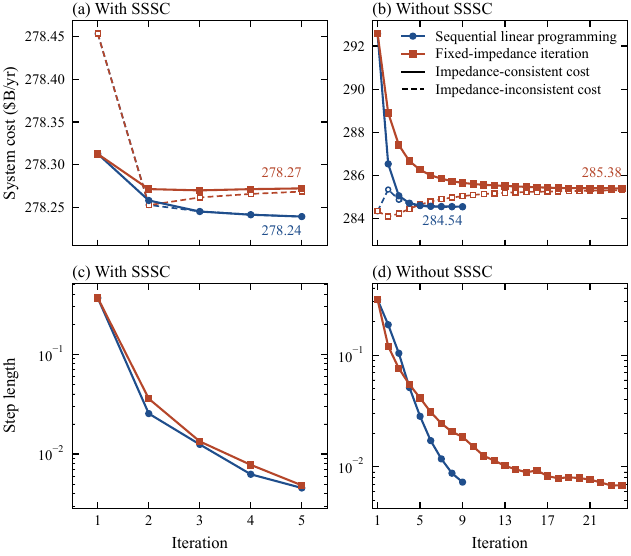}
\caption{Convergence of SLP and FI on a demonstration case with and without SSSCs. (a)--(b) System cost. Solid lines: impedance-consistent cost, obtained by fixing the AC line capacities at the current iterate and re-optimizing. Dashed lines: impedance-inconsistent cost reported by the subproblem. (c)--(d) Step length $\|\mathbf{F}^{n}-\mathbf{F}^{n-1}\|_1/\|\mathbf{F}^0\|_1$.}
\label{fig:convergence}
\end{figure}

Figure~\ref{fig:convergence} shows a substantial gap between the impedance-inconsistent and impedance-consistent costs before convergence, indicating that planning results cannot be reliably assessed without impedance consistency. SLP converges faster and reaches a lower system cost than FI, especially without SSSCs, where it converges in 9 rather than 24 iterations and attains 284.54 rather than 285.38~\$B/yr. These results demonstrate the advantages of SLP over FI and are consistent with the optimality analysis in Section~\ref{sec:kkt}. Although full SLP convergence requires 5 to 9 iterations, 2 to 3 iterations capture most of the reduction in impedance-consistent cost, so stopping SLP after 2–3 iterations is a practical option when computational resources are limited for transmission planning with or without SSSCs.

\subsection{Least-Cost Expansion Plan and SSSC Deployment}
\label{sec:res_plan}

\begin{figure}[t]
\centering
\includegraphics[width=\columnwidth]{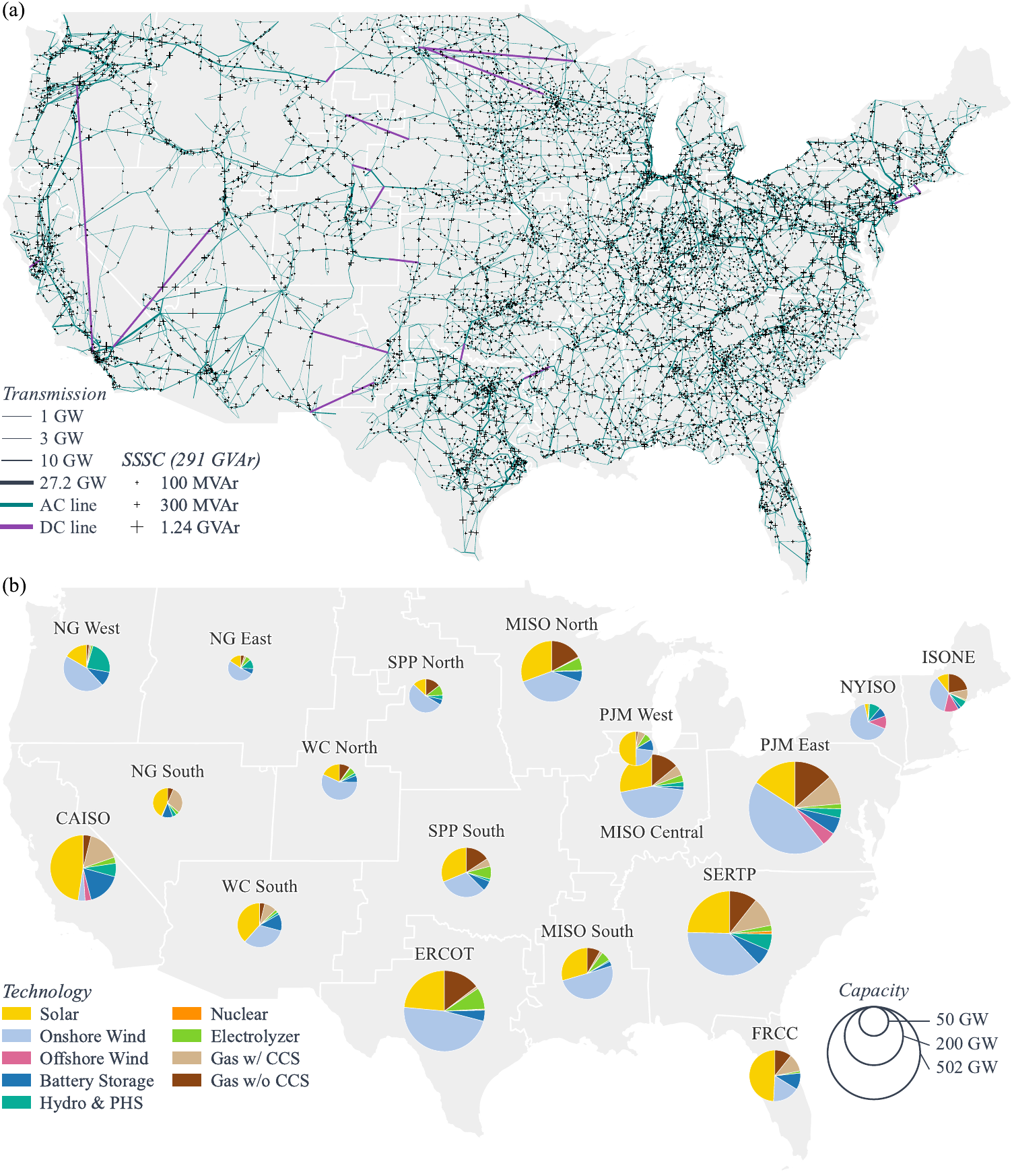}
\caption{Least-cost expansion plan in the baseline scenario. (a) Transmission and SSSC deployment. SSSC deployments are shown as + markers at the midpoint of AC lines. (b) Total installed generation, storage, and electrolyzer capacities by planning sub-region.}
\label{fig:map}
\end{figure}

Figure~\ref{fig:map} presents the least-cost expansion plan in the baseline scenario. SSSCs are widely deployed across the country, with a total capacity of 291~GVAr. Wind and solar account for the majority of installed capacity in most regions, complemented by thermal generation, battery storage, hydro and pumped storage, and electrolyzers.

\begin{figure}[t]
\centering
\includegraphics[width=\columnwidth]{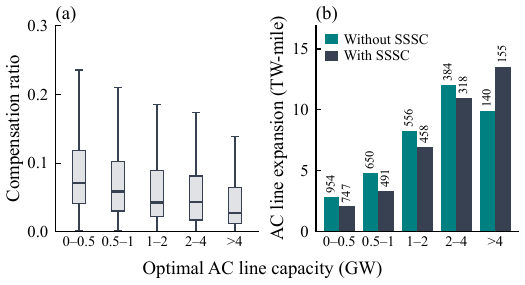}
\caption{(a) Distribution of SSSC compensation ratio $Q_{\mathrm{SSSC},\ell}/F_\ell$, the SSSC rating installed on a line relative to the line's capacity, by optimal AC line capacity interval. (b) AC line expansion volume by optimal AC line capacity interval with and without SSSCs. Numbers above the bars indicate the number of AC lines expanded by more than 50~MW. Baseline scenario.}
\label{fig:compensation}
\end{figure}

Figure~\ref{fig:compensation}(a) shows the compensation ratio $Q_{\mathrm{SSSC},\ell}/F_\ell$, the SSSC rating installed on a line relative to the line's capacity, across optimal AC line capacity intervals. The median compensation ratio decreases from approximately 0.07 on lines below 0.5~GW to approximately 0.02 on lines above 4~GW. This pattern reflects compensation feedback: because the SSSC term in~\eqref{eq:kvl} scales with $Q_{\mathrm{SSSC},\ell}/F_\ell$, each unit of SSSC rating provides more compensation on a smaller line.

Figure~\ref{fig:compensation}(b) compares AC line expansion volume across optimal line capacity intervals. With SSSCs, expansion decreases on lines below 4~GW and increases on lines above 4~GW, which typically serve as backbone lines. Flow control relieves congestion on small-to-medium lines, allowing the binding transmission expansion budget to be concentrated on higher-capacity lines. From without to with SSSCs, the number of AC lines expanded by more than 50~MW decreases from 2,684 to 2,169, a reduction of 19\%, with only 15 more lines being expanded in the highest-capacity interval.

\subsection{Economic Value of SSSCs}
\label{sec:res_value}

\begin{figure}[t]
\centering
\includegraphics[width=\columnwidth]{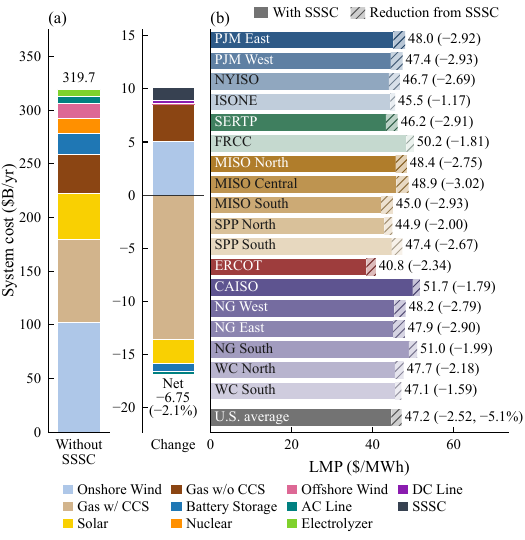}
\caption{Baseline system value of SSSCs. (a) Annualized cost without SSSCs and changes by component. (b) Average regional and national LMPs with SSSCs; hatched segments and parenthetical values indicate reductions from deployment.}
\label{fig:cost_lmp}
\end{figure}

Figure~\ref{fig:cost_lmp}(a) decomposes the annualized system cost and its change induced by SSSC deployment. Without SSSCs, the baseline cost is 319.7~\$B/yr. SSSCs reduce it by 6.75~\$B/yr (2.1\%). We define the overall benefit--cost ratio (BCR) as the annualized reduction in all system costs other than SSSC costs divided by the annualized SSSC cost; it equals 6.4 in the baseline. SSSC power flow control enables greater onshore wind integration, reducing reliance on costly gas-fired generation with CCS for firm capacity and yielding the largest component saving. This saving is partly offset by more investment in onshore wind for energy and unabated gas used for peaking capacity. Solar, battery storage, and AC line costs also decrease. These portfolio shifts show that the value of SSSCs extends beyond transmission investment to the generation and storage portfolio.

Figure~\ref{fig:cost_lmp}(b) shows lower average locational marginal prices (LMPs) in all 18 regions, indicating that no region is disadvantaged by loop-flow control in terms of average wholesale electricity prices. The national average with SSSCs is 47.2~\$/MWh, a reduction of 2.52~\$/MWh (5.1\%). Regional reductions range from 1.17~\$/MWh in ISONE to 3.02~\$/MWh in MISO Central. The relative wholesale price reduction far exceeds the relative system cost saving and indicates a potential transfer of surplus from generators to consumers.

\begin{figure}[t]
\centering
\includegraphics[width=0.88\columnwidth]{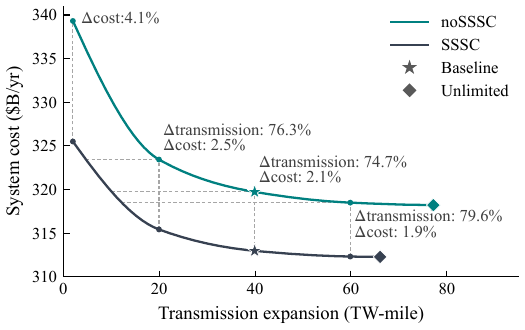}
\caption{Optimal system cost versus transmission expansion volume. Baseline: total transmission expansion limited to 20\% of the current network volume; Unlimited: transmission expansion unlimited. $\Delta$transmission indicates transmission expansion savings at the same system cost. $\Delta$cost indicates cost savings from SSSCs at each transmission expansion limit.}
\label{fig:tx_limit}
\end{figure}

To examine the interaction with transmission expansion, we vary the total transmission expansion limit~\eqref{eq:trans_limit} and measure expansion volume as added capacity times line length in TW-mile (Fig.~\ref{fig:tx_limit}). Vertical comparisons between the curves quantify the cost savings from SSSCs at each expansion limit. These relative savings grow as the limit becomes more stringent: 1.9\% with unlimited expansion, 2.1\% at the baseline limit of 20\% of the current network volume, and 4.1\% at the tightest tested limit of 1\%. SSSCs are therefore most valuable where transmission expansion is most constrained. Horizontal comparisons instead measure how much transmission expansion SSSCs could avoid without increasing system cost. Holding system cost at the no-SSSC baseline level of 319.7~\$B/yr, SSSCs reduce the required expansion from approximately 40 to 10.1~TW-mile, avoiding 74.7\% of it.  At the cost levels of the no-SSSC plans with 20~TW-mile and unlimited transmission expansion, SSSCs avoid 76.3\% and 79.6\% of the transmission expansion, respectively. Thus, SSSCs could avoid roughly three-quarters of conventional transmission expansion without increasing system cost.

\subsection{High-Value SSSC Deployment Opportunities}
\label{sec:res_highvalue}

\begin{figure}[t]
\centering
\includegraphics[width=\columnwidth]{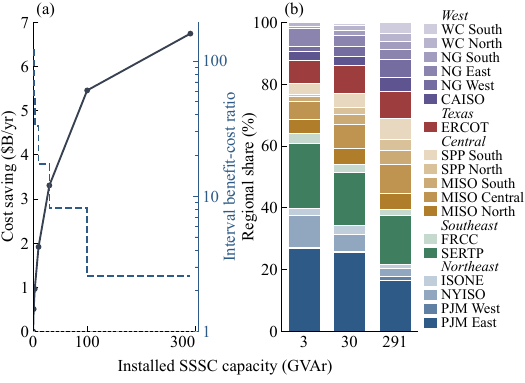}
\caption{benefit--cost ratio and regional allocation of SSSC deployment in the baseline. (a) Cost savings (left axis) and interval BCR (right axis). (b) Regional SSSC capacity shares at 3, 30, and 291~GVAr.}
\label{fig:marginal}
\end{figure}

To identify the high-value deployment opportunities, we impose total SSSC capacity limits of 1, 3, 10, 30, and 100~GVAr and compare them with unrestricted deployment of 291~GVAr. Figure~\ref{fig:marginal}(a) reports cost savings and the interval BCRs, defined as the incremental cost reduction divided by the incremental SSSC cost. The interval BCR exceeds 100 for the first 1~GVAr and declines as deployment expands. Notably, only one-tenth of the optimal capacity (30~GVAr) captures nearly half of the cost savings; 100~GVAr, approximately one-third, captures about 80\%. Thus, a small fraction of the optimal capacity secures a large share of the value.

Figure~\ref{fig:marginal}(b) locates these high-value opportunities. At 3~GVAr, PJM East and SERTP together account for half of national deployment, and approximately 80\% is in the Eastern Interconnection. As SSSC deployment expands, it covers opportunities with relatively lower BCRs across more regions. The shares of PJM East and SERTP gradually decrease, while those of the Western and Central regions increase. These patterns support prioritizing eastern opportunities initially and broadening deployment as the highest-return investments are exhausted.

\subsection{Sensitivity Analysis}
\label{sec:res_sensitivity}

\begin{figure}[t]
\centering
\includegraphics[width=1\columnwidth]{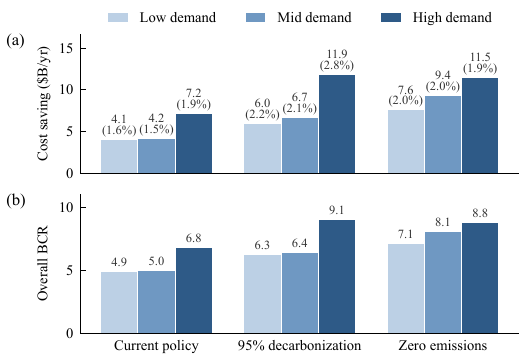}
\caption{Sensitivity to demand and decarbonization. Low, Mid, and High demand: approximately 5,500, 6,400, and 8,000~TWh/yr. Current policy, 95\% decarbonization, and Zero emissions: 0, 476, and 1512~TWh/yr of electrolysis load. (a) Cost savings, with relative savings in parentheses. (b) Overall BCR.}
\label{fig:sensitivity}
\end{figure}

Figure~\ref{fig:sensitivity} illustrates the sensitivity of SSSC value to demand and decarbonization scenarios. In addition to the ``Mid'' demand of the baseline, we consider ``Low'' and ``High'' demand scenarios with annual exogenous demand of approximately 5,500 and 8,000~TWh, respectively. In addition to the ``95\% decarbonization'' of the baseline, we consider a ``Current policy'' scenario, which retains only the current policies described in Section~\ref{sec:setup}, and a ``Zero emissions'' scenario, which mandates 100\% zero-carbon generation. The flexible electrolysis load increases with the stringency of decarbonization, from 0~TWh under ``Current policy'' to 476~TWh under ``95\% decarbonization'' and 1,512~TWh under ``Zero emissions'' \cite{EvolvedEnergy_ADP2024}. Across all nine scenarios, SSSCs yield annualized cost savings of 4.1--11.9~\$B/yr (1.5--2.8\%) and overall BCRs of 4.9--9.1. Demand growth increases absolute savings and BCR within each decarbonization scenario. Both decarbonization scenarios also provide higher savings and BCRs than Current policy, consistent with greater value from delivering low-cost renewable energy across the network.

Finally, we double the SSSC cost to 120~\$/kVAr (Table~\ref{tab:comparison}). This large cost increase has a relatively small effect on system value: annualized savings are still 5.7~\$B/yr, only 15.6\% below the baseline, while the overall BCR is still 4.1. Installed capacity falls to 213~GVAr, a reduction of 26.8\%. The system therefore retains approximately 84\% of the baseline savings despite a greater reduction in deployment.

\begin{table}[t]
\centering
\begin{threeparttable}
\caption{Sensitivity of SSSC Value to SSSC Cost\label{tab:comparison}}
\begin{tabular}{@{}lccc@{}}
\toprule
Scenario & \makecell{Cost savings} & \makecell{SSSC capacity} & \makecell{Overall BCR} \\
\midrule
Baseline & 6.75 \$B/yr & 291 GVAr & 6.4 \\
High SSSC cost (2$\times$) & 5.69 \$B/yr & 213 GVAr & 4.1 \\
\bottomrule
\end{tabular}
\end{threeparttable}
\end{table}

\section{Discussion and Conclusion}
\label{sec:conclusion}
This paper develops a capacity expansion model of the contiguous U.S. power system for 2050 that co-optimizes generation, storage, electrolysis, transmission, and SSSCs, and applies it across transmission expansion, demand, decarbonization, and device cost scenarios. The model captures transmission feedback, which makes the problem nonconvex: expanding a line lowers its impedance (impedance feedback) and the series compensation its SSSC can provide (compensation feedback). We show that converged solutions of the conventional FI iteration violate stationarity because the iteration does not internalize transmission feedback. SLP with adaptive proximal regularization internalizes it and reaches a KKT point. In the numerical tests, it also converges in fewer iterations to lower-cost plans.

In the baseline scenario, SSSCs reduce annualized system cost by \$6.75~billion (2.1\%) with an overall BCR of 6.4 and lower the national average LMP by 5.1\%. Alternatively, they could avoid about three-quarters of the baseline transmission expansion without increasing system cost. Across demand and decarbonization scenarios, overall BCRs of SSSCs range from 4.9 to 9.1, and remain 4.1 when the device cost is doubled. Flow control on small-to-medium lines concentrates expansion on high-capacity lines and reduces the number of substantially expanded lines by 19\%. Because the model prices transmission expansion linearly, accounting for its economies of scale would likely increase the system cost savings. Fast flow control could also reduce redispatch costs under forecast errors and contingencies, a benefit not quantified here. The deployment results support prioritizing the highest-BCR opportunities before pursuing the full optimal portfolio. The first 1~GVAr of deployment has an interval BCR above 100 and is concentrated in eastern load centers, especially PJM East and SERTP. These regions are promising locations for early SSSC demonstrations, which could build operational experience and support technological improvement through learning by doing.

\bibliographystyle{IEEEtran}
\bibliography{ref}

@misc{FERC_Staff_Reports,
  author       = {{Federal Energy Regulatory Commission}},
  title        = {Staff Reports and Papers},
  url          = {https://www.ferc.gov/staff-reports-and-papers},
}

@misc{HIFLD,
  author       = {{Homeland Infrastructure Foundation-Level Data (HIFLD)}},
  title        = {{U.S.} Electric Power Transmission Lines},
  url          = {https://www.arcgis.com/home/item.html?id=d4090758322c4d32a4cd002ffaa0aa12},
  year    = {2024},
}

@article{Mai2026,
  author  = {Mai, Trieu and Brown, Patrick R. and Palchak, David and Brinkman, Gregory and Brooks, Adria and Denholm, Paul and Gramlich, Rob and Homer, Juliet S. and Lew, Debra and Kuna, Jess and Okullo, James and Pfeifenberger, Johannes and Rose, Amy and Simeone, Christina and Wright, Jarrad},
  title   = {What Can We Learn from {US} National Transmission Studies?},
  journal = {Joule},
  pages   = {102339},
  year    = {2026},
  doi     = {10.1016/j.joule.2026.102339},
}

@techreport{DOE_ATT_2020,
  author      = {{U.S. Department of Energy}},
  title       = {Advanced Transmission Technologies},
  institution = {U.S. Department of Energy},
  year        = {2020},
  month       = dec
}

@article{RN1348,
   author = {Hörsch, Jonas and Ronellenfitsch, Henrik and Witthaut, Dirk and Brown, Tom},
   title = {Linear optimal power flow using cycle flows},
   journal = {Electric Power Systems Research},
   volume = {158},
   pages = {126-135},
   ISSN = {03787796},
   DOI = {10.1016/j.epsr.2017.12.034},
   year = {2018},
   type = {Journal Article}
}

@book{Diestel2025,
  author    = {Reinhard Diestel},
  title     = {Graph Theory},
  edition   = {6},
  year      = {2025},
  publisher = {Springer},
  series    = {Graduate Texts in Mathematics},
  address   = {Berlin, Heidelberg},
}

@article{RN1350,
   author = {Alajrash, Ban H. and Salem, Mohamed and Swadi, Mahmood and Senjyu, Tomonobu and Kamarol, Mohamad and Motahhir, Saad},
   title = {A comprehensive review of {FACTS} devices in modern power systems: Addressing power quality, optimal placement, and stability with renewable energy penetration},
   journal = {Energy Reports},
   volume = {11},
   pages = {5350-5371},
   ISSN = {23524847},
   DOI = {10.1016/j.egyr.2024.05.011},
   year = {2024},
   type = {Journal Article}
}

@article{Wu2024EPSR,
   author = {Wu, Kevin and Tanneau, Mathieu and Hentenryck, Pascal Van},
   title = {Strong mixed-integer formulations for transmission expansion planning with {FACTS} devices},
   journal = {Electric Power Systems Research},
   volume = {235},
   ISSN = {03787796},
   DOI = {10.1016/j.epsr.2024.110695},
   year = {2024},
   type = {Journal Article}
}

@article{Franken2020IET,
   author = {Franken, Marco and Barrios, Hans and Schrief, Alexander B. and Moser, Albert},
   title = {Transmission expansion planning via power flow controlling technologies},
   journal = {IET Generation, Transmission \& Distribution},
   volume = {14},
   number = {17},
   pages = {3530-3538},
   ISSN = {1751-8687
1751-8695},
   DOI = {10.1049/iet-gtd.2019.1897},
   year = {2020},
   type = {Journal Article}
}

@misc{ENTSOE_SSSC_Technopedia,
  author       = {{ENTSO-E}},
  title        = {Static Synchronous Series Compensator ({SSSC})},
  year         = {2025},
  url          = {https://www.tsodsoplatform.eu/technopedia/factsheets/static-synchronous-series-compensators-sssc},
}

@incollection{nilsson2020application,
  title={Application Examples of {UPFC} and Its Variants},
  author={Nilsson, Stig L. and Xu, Shukai and Lei, Bo and Deng, Zhanfeng and Andersen, Bjarne R.},
  editor={Andersen, Bjarne R. and Nilsson, Stig L.},
  booktitle={Flexible {AC} Transmission Systems: {FACTS}},
  series={CIGRE Green Books},
  pages={645--706},
  year={2020},
  publisher={Springer},
  address={Cham},
  doi={10.1007/978-3-030-35386-5_15}
}

@techreport{EPRI_APFC_2024,
  title        = {Advanced Power Flow Controllers ({APFC})},
  author  = {{EPRI}},
  institution  = {Electric Power Research Institute},
  year         = {2024},
  url          = {https://www.epri.com/research/products/000000003002030548},
}

@article{Lee2026TPWRS,
   author = {Lee, Thomas and Sun, Andy},
   title = {Contingency-Aware Nodal Optimal Power Investments with High Temporal Resolution},
   journal = {IEEE Transactions on Power Systems},
   pages = {1-13},
   ISSN = {0885-8950
1558-0679},
   DOI = {10.1109/tpwrs.2026.3739686},
   year = {2026},
   type = {Journal Article}
}

@article{Zhang2017TPWRS,
   author = {Zhang, Xiaohu and Tomsovic, Kevin and Dimitrovski, Aleksandar},
   title = {Security Constrained Multi-Stage Transmission Expansion Planning Considering a Continuously Variable Series Reactor},
   journal = {IEEE Transactions on Power Systems},
   volume = {32},
   number = {6},
   pages = {4442-4450},
   ISSN = {0885-8950
1558-0679},
   DOI = {10.1109/tpwrs.2017.2671786},
   year = {2017},
   type = {Journal Article}
}

@misc{smartwires2019,
  author       = {{Smart Wires, Inc.}},
  title        = {Comments on the 2019--2020 Transmission Planning Process},
  howpublished = {Comments submitted to the California Independent System Operator (CAISO)},
  year         = {2019},
  month        = dec,
  url          = {https://www.caiso.com/Documents/SmartWiresComments-2019-2020TransmissionPlanningProcess-Nov182019Meeting-SubmittedDec262019.pdf}
}

@article{HAGSPIEL2014654,
title = {Cost-optimal power system extension under flow-based market coupling},
journal = {Energy},
volume = {66},
pages = {654-666},
year = {2014},
issn = {0360-5442},
doi = {https://doi.org/10.1016/j.energy.2014.01.025},
author = {S. Hagspiel and C. Jägemann and D. Lindenberger and T. Brown and S. Cherevatskiy and E. Tröster},
}

@techreport{EvolvedEnergy_ADP2024,
  title        = {{Annual Decarbonization Perspective} 2024},
  author       = {{Evolved Energy Research}},
  institution  = {Evolved Energy Research},
  year         = {2024},
  url          = {https://www.evolved.energy/us-adp-2024},
}

@misc{NREL_ReEDS,
  title        = {Model documentation — ReEDS 2.0},
  author       = {{National Laboratory of the Rockies}},
  year         = {2026},
  howpublished = {\url{https://reeds-model.github.io/ReEDS/model_documentation.html}},
}

@misc{Tehranchi_PyPSAUSA2024,
  title        = {{PyPSA-USA}: A Flexible Open-Source Energy System Model and Optimization Tool for the {United States}},
  author       = {Tehranchi, Kamran and Barnes, Trevor and Frysztacki, Martha and Azevedo, In{\^e}s M. L.},
  year         = {2024},
  howpublished = {SSRN preprint},
  doi          = {10.2139/ssrn.5029120},
  url          = {https://doi.org/10.2139/ssrn.5029120}
}

@article{Brown_PyPSA2018,
  title   = {{PyPSA}: {Python} for Power System Analysis},
  author  = {Brown, Tom and H{\"o}rsch, Jonas and Schlachtberger, David},
  year    = {2018},
  journal = {Journal of Open Research Software},
  volume  = {6},
  number  = {1},
  pages   = {4},
  doi     = {10.5334/jors.188},
  url     = {https://doi.org/10.5334/jors.188}
}

@article{Li2018TPWRS,
   author = {Li, Jia and Li, Zuyi and Liu, Feng and Ye, Hongxing and Zhang, Xuemin and Mei, Shengwei and Chang, Naichao},
   title = {Robust Coordinated Transmission and Generation Expansion Planning Considering Ramping Requirements and Construction Periods},
   journal = {IEEE Transactions on Power Systems},
   volume = {33},
   number = {1},
   pages = {268-280},
   ISSN = {0885-8950
1558-0679},
   DOI = {10.1109/tpwrs.2017.2687318},
   year = {2018},
   type = {Journal Article}
}

@article{Hamidpour2021IJEPES,
   author = {Hamidpour, Hamidreza and Pirouzi, Sasan and Safaee, Sheila and Norouzi, Mohammadali and Lehtonen, Matti},
   title = {Multi-objective resilient-constrained generation and transmission expansion planning against natural disasters},
   journal = {International Journal of Electrical Power \& Energy Systems},
   volume = {132},
   ISSN = {01420615},
   DOI = {10.1016/j.ijepes.2021.107193},
   year = {2021},
   type = {Journal Article}
}

@article{Mokhtari2021EPSR,
   author = {Mokhtari, Mohammad Sadegh and Gitizadeh, Mohsen and Lehtonen, Matti},
   title = {Optimal coordination of thyristor controlled series compensation and transmission expansion planning: Distributionally robust optimization approach},
   journal = {Electric Power Systems Research},
   volume = {196},
   ISSN = {03787796},
   DOI = {10.1016/j.epsr.2021.107189},
   year = {2021},
   type = {Journal Article}
}

@misc{Selvans2026_PUDL,
  author       = {Selvans, Zane A. and Gosnell, Christina M. and Sharpe, Austen and Schira, Zach and Xia, Dazhong and Belfer, Ella and Mazaitis, Kathryn},
  title        = {Public Utility Data Liberation Project {(PUDL)} Data Release},
  year         = {2026},
  version      = {v2026.5.0},
  doi          = {10.5281/zenodo.20275549},
  url          = {https://doi.org/10.5281/zenodo.20275549},
}

@techreport{NLR_ATB_2025,
  author      = {{NLR}},
  title       = {2025 {Annual Technology Baseline}},
  institution = {National Laboratory of the Rockies},
  year        = {2026},
  url         = {https://atb.nlr.gov/},
}

@inproceedings{xu2020us,
  title={{U.S.} Test System with High Spatial and Temporal Resolution for Renewable Integration Studies},
  author={Xu, Yixing and Myhrvold, Nathan and Sivam, Dhileep and Mueller, Kaspar and Olsen, Daniel J. and Xia, Bainan and Livengood, Daniel and Hunt, Victoria and d'Orfeuil, Benjamin Rouill{\'e} and Muldrew, Daniel and Ondreicka, Merrielle and Bettilyon, Megan},
  booktitle={2020 IEEE Power \& Energy Society General Meeting (PESGM)},
  pages={1--5},
  year={2020},
  doi={10.1109/PESGM41954.2020.9281850}
}

@article{RN1333,
   author = {Mirzapour, Omid and Rui, Xinyang and Sahraei-Ardakani, Mostafa},
   title = {Grid-enhancing technologies: Progress, challenges, and future research directions},
   journal = {Electric Power Systems Research},
   volume = {230},
   ISSN = {03787796},
   DOI = {10.1016/j.epsr.2024.110304},
   year = {2024},
   type = {Journal Article}
}

@article{Singh_EPSR_2021,
   author = {Singh, Pradeep and Senroy, Nilanjan},
   title = {Steady-State models of {STATCOM} and {UPFC} using flexible holomorphic embedding},
   journal = {Electric Power Systems Research},
   volume = {199},
   ISSN = {03787796},
   DOI = {10.1016/j.epsr.2021.107390},
   year = {2021},
   type = {Journal Article}
}

@article{gazafroudi2022topology,
  title={Topology-based approximations for {$N-1$} contingency constraints in power transmission networks},
  author={Shokri Gazafroudi, Amin and Neumann, Fabian and Brown, Tom},
  journal={International Journal of Electrical Power \& Energy Systems},
  volume={137},
  pages={107702},
  year={2022},
  publisher={Elsevier},
  doi={10.1016/j.ijepes.2021.107702}
}

@misc{ai_delineating,
  title        = {Delineating the Roles of Hydrogen-Based Flexibility Pathways},
  author       = {Ai, Wei and Dowling, Jacqueline A. and Craig, Michael T.},
  year         = {2026},
  howpublished = {SSRN preprint},
  doi          = {10.2139/ssrn.7321278},
  url          = {https://doi.org/10.2139/ssrn.7321278},
}

@article{Spielman2011,
   author = {Spielman, Daniel A. and Srivastava, Nikhil},
   title = {Graph Sparsification by Effective Resistances},
   journal = {SIAM Journal on Computing},
   volume = {40},
   number = {6},
   pages = {1913-1926},
   ISSN = {0097-5397
1095-7111},
   DOI = {10.1137/080734029},
   year = {2011},
   type = {Journal Article}
}

@article{Grudzien2018,
   author = {Grudzien, Colin and Deka, Deepjyoti and Chertkov, Michael and Backhaus, Scott N.},
   title = {Structure- and Physics-Preserving Reductions of Power Grid Models},
   journal = {Multiscale Modeling \& Simulation},
   volume = {16},
   number = {4},
   pages = {1916-1947},
   ISSN = {1540-3459
1540-3467},
   DOI = {10.1137/17m1138054},
   year = {2018},
   type = {Journal Article}
}

@misc{FERC_Order1920,
  author       = {{Federal Energy Regulatory Commission}},
  title        = {Building for the Future Through Electric Regional Transmission Planning and Cost Allocation, {Order No.~1920}, 187 {FERC} {\P}~61,068},
  year         = {2024},
  month        = may,
}

@misc{DOE_SPARK2026,
  author       = {{U.S. Department of Energy}},
  title        = {{Energy Department} Announces {Speed to Power} Investments Across 26 States to Lower Electricity Costs and Improve Grid Reliability},
  year         = {2026},
  month        = sep,
  url          = {https://www.energy.gov/articles/energy-department-announces-speed-power-investments-across-26-states-lower-electricity},
}

@article{palacios1982nonlinear,
  title={Nonlinear optimization by successive linear programming},
  author={Palacios-Gomez, Fernando and Lasdon, L. and Engquist, Michael},
  journal={Management Science},
  volume={28},
  number={10},
  pages={1106--1120},
  year={1982},
  publisher={INFORMS},
  doi={10.1287/mnsc.28.10.1106}
}

@article{Divan2007DSSSC,
   author = {Divan, Deepak M. and Brumsickle, William E. and Schneider, Robert S. and Kranz, Bill and Gascoigne, Randal W. and Bradshaw, Dale T. and Ingram, Michael R. and Grant, Ian S.},
   title = {A Distributed Static Series Compensator System for Realizing Active Power Flow Control on Existing Power Lines},
   journal = {IEEE Transactions on Power Delivery},
   volume = {22},
   number = {1},
   pages = {642-649},
   ISSN = {0885-8977},
   DOI = {10.1109/tpwrd.2006.887103},
   year = {2007},
   type = {Journal Article}
}

\end{document}